\documentclass[prl,twocolumn,showpacs]{revtex4}
\usepackage{graphicx}
\usepackage{bm}
\usepackage{subfigure}
\usepackage{amsmath}
\usepackage{color}

\newcommand{\pr}{Phys. Rev.\ }

\newcommand{\etal}{{\em et al.\ }}

\newcommand{\UQ}{School of Mathematics and Physics, University of Queensland, Brisbane, 
QLD 4072, Australia.}
\begin{document}

\title{Comment on ``Negative Differential Conductivity in an Interacting Quantum Gas.'' }

\author{M.~K. Olsen and J.~F. Corney}
\affiliation{\UQ}
%-----------------------------------------------------------------------
\date{\today}
%------------------------------------------------------------------------

\begin{abstract}

Labouvie \etal (\prl {\bf 115}, 050601, (20015)) recently  demonstrated negative differential conductivity (NDC)  in a multi-well Bose-Einstein condensate. They stated ``we demonstrate that NDC originates from a nonlinear, atom number dependent tunneling coupling in combination with fast collisional decoherence.'' We show theoretically how the essential feature of NDC, a reduction in atomic current caused by an increase in chemical potential, is present in unitary dynamics through the well-known mechanism of macroscopic self-trapping (MST), and that the collisional decoherence merely serves as a quantitative modification of this.

\end{abstract}
%******************************************* 


\maketitle

NDC is an unusual phenomenon in electronics which requires a strongly nonlinear device.  In ultracold-atom transport, nonlinearity is readily available through atomic collisions.  An atomic implementation of NDC by Labouvie \etal~\cite{NDC} found that, if chemical potential difference, $\Delta\mu$, between wells is considered as analogous to voltage, the proportionality between $\Delta\mu$ and tunneling current can be negative. We show here how the qualitative effects of this phenomenon can be ascribed to the well known MST phenomenon (MST)~\cite{BHJoel}, with the collisional decoherence causing only quantitative changes. 
This means that NDC is available in a lattice ultracold atomic system purely through coherent effects.

We analyse a three-well Bose-Hubbard model~\cite{BHmodel}, described by the Hamiltonian
\begin{equation}
{\cal H} = \hbar\chi\sum_{i=1}^{3}\hat{a}_{i}^{\dag\,2}\hat{a}_{i}^{2}-\hbar J\left( \hat{a}_{1}^{\dag}\hat{a}_{2}+\hat{a}_{2}^{\dag}\hat{a}_{1}+\hat{a}_{3}^{\dag}\hat{a}_{2}+\hat{a}_{2}^{\dag}\hat{a}_{3} \right),
\label{eq:Ham}
\end{equation} 
using the truncated Wigner representation~\cite{Steel,Chiancathermal}. $\chi$ is the collisional nonlinearity, $J$ is the tunneling parameter, and phase diffusion is included by the same Louivillian proportional to  $\Gamma$ as used by Labouvie \etal
We define $\Delta N\equiv N_{1}(0)-N_{2}(0)$, approximately proportional to the difference in chemical potential between the wells and analogous to voltage since it  drives the atomic current.

\begin{figure}[tbhp]
\begin{center}
\includegraphics[width=0.75\columnwidth]{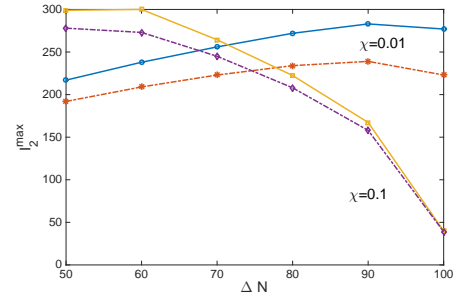}  
\end{center}
\caption{(Colour online) The  maximum currents into the middle well as a function of $\Delta N$, for $J=1$, $\chi=0.01$ and $0.1$, and $N_{1}(0)=N_{3}(0)=100$, with $N_{2}(0)=N_{1}(0)-\Delta N$. The solid lines are the results for initial coherent states with $\Gamma_{2}=0$ while the dashed lines include phase diffusion with $\Gamma_{2}=1.5J$. Each result is the average of the order of a million stochastic trajectories and sampling errors are within line thicknesses.}
\label{fig:current1}
\end{figure}

We calculate the maximum of the tunnelling current, $I_2 = -i\langle \hat{a}_{2}^{\dag}\hat{a}_{1}-\hat{a}_{1}^{\dag}\hat{a}_{2}+\hat{a}_{3}^{\dag}\hat{a}_{2}-\hat{a}_{2}^{\dag}\hat{a}_{2}\rangle$,
for different $\Delta N$ and two values of $\chi$, with initial coherent states~\cite{states}.
The results are shown in Fig.~\ref{fig:current1}, with ($\Gamma=1.5J$) and without phase noise in the central well. The stronger nonlinearity leads to a decrease of current with increasing $\Delta N$, which is an example of NDC.  We see that the presence of phase noise does not qualitatively change the results.

These numerical results show that phase diffusion is not necessary for negative differential conductivity, which can arise solely from macroscopic self-trapping, a fully coherent process. We note here that the direct current reported by Labouvie \etal does depend on the presence of phase diffusion, but does not exist for the parameters we consider here. In our view, the new effect that has been discovered is this DC manifestation of NDC, and not NDC itself.

This research was supported by the Australian Research Council under the Future Fellowships Program (Grant ID: FT100100515).

\end{document}